\documentstyle[11pt,aasms4,overcite]{article}

\def\gsim{\;\rlap{\lower 2.5pt
\hbox{$\sim$}}\raise 1.5pt\hbox{$>$}\;}
\def\lsim{\;\rlap{\lower 2.5pt
   \hbox{$\sim$}}\raise 1.5pt\hbox{$<$}\;}

\def\ge{\;\rlap{\lower 2.5pt
 \hbox{$-$}}\raise 1.5pt\hbox{$>$}\;}
\def\le{\;\rlap{\lower 2.5pt
   \hbox{$-$}}\raise 1.5pt\hbox{$<$}\;}

\newcommand\beq{\begin{equation}} 
\newcommand\eeq{\end{equation}}

\begin{document}

\title{Photometric Variability in Earthshine Observations}

\author{Sally V. Langford\altaffilmark{1,2}, J. Stuart B. Wyithe\altaffilmark{1} and 
Edwin L. Turner\altaffilmark{2,3}}

\altaffiltext{1}{School of Physics, University of Melbourne, Parkville, Victoria, Australia.}
\altaffiltext{2}{Princeton University Observatory, Princeton, New Jersey.}
\altaffiltext{3}{Institute for the Physics and Mathematics of the Universe, University of Tokyo, Kashiwa, Japan}

{\bf The identification of an extrasolar planet as Earth-like will
  depend on the detection of atmospheric signatures or surface
  non-uniformities.  In this paper we present spatially unresolved
  flux light curves of Earth for the purpose of studying a prototype
  extrasolar terrestrial planet.  Our monitoring of the photometric
  variability of earthshine revealed changes of up to $23 \%$ per hour
  in the brightness of Earth's scattered light at around 600 nm, due
  to the removal of specular reflection from the view of the Moon.
  This variability is accompanied by reddening of the spectrum, and
  results from a change in surface properties across the continental
  boundary between the Indian Ocean and Africa's east coast.  Our
  results based on earthshine monitoring indicate that specular
  reflection should provide a useful tool in determining the presence
  of liquid water on extrasolar planets via photometric observations.
}


\subsection*{Introduction}
A primary goal of future space-based missions, such as the European
Space Agency's Darwin and the National Aeronautics and Space
Administration's Terrestrial Planet Finder (TPF), is the discovery of
planets in the habitable zones of nearby stars in the hopes of
detecting an Earth-like planet (Des Marais~{\it et~al.}, 2001; Howard
and Horowitz, 2001).  Rapid, large amplitude variation in the
photometric flux of a planet is thought to be uniquely terrestrial;
indeed, modeling predicts daily variations of around $20 \%$ in
Earth's spatially unresolved visible light curve (Ford~{\it et~al.},
2001).  Daily variations are predicted to be greater for a cloud-free
Earth (Tinetti~{\it et~al.}, 2006).  This variation depends on the
changing surface properties and weather conditions on Earth as the
planet rotates (Ford~{\it et~al.}, 2001).  Large cloud formations are
generally tied to surface features but dissipate on the time scale of
days, which allows the long term rotational period of Earth to be
discernible from folded light curves (Pall\'{e}~{\it et~al.}, 2008). Thus, the
study of the photometric variability of Earth (for which we have
detailed surface knowledge) will be directly relevant to interpreting
observations of extrasolar terrestrial planets (Seager~{\it et~al.}, 2003).

Sunlight scattered by Earth illuminates the side of the Moon facing
away from the Sun (known as earthshine) and permits measurement of the
spatially unresolved (total) brightness of the Earth.  Therefore,
earthshine provides a view of Earth that is analogous to observations
of an extrasolar planet (Seager~{\it et~al.}, 2003).  A number of
authors have observed earthshine for the purposes of climate
monitoring and exploration of the possible features of the disk
averaged spectrum (Goode~{\it et~al.}, 2001; Woolf~{\it et~al.}, 2002;
Hamdani~{\it et~al.}, 2006; Monta\~{n}\'{e}s-Rodr\'{i}guez~{\it
  et~al.}, 2006; Turnbull~{\it et~al.}, 2006; Arnold, 2007); it is
predicted that other important properties may be determined from
intra-day photometric variability (Ford~{\it et~al.}, 2001; Seager, 2003;
Pall\'{e}~{\it et~al.}, 2008).  Earthshine is dominated by a relatively small
region of the hemisphere visible from the Moon due to the viewing
angle (Des Marais~{\it et~al.}, 2002; Seager~{\it et~al.}, 2005; Pall\'{e}~{\it et~al.},
2008).  Fresnel (specular) reflection (at a point determined by the
equality of the angles of incidence and reflection) dominates in
oceanic regions, which otherwise have a low albedo (Sagan~{\it et~al.},
1993). However rotation of land (which scatters light diffusively)
onto the location of this bright spot interrupts the specular
reflection, which leads to a predicted decrease in earthshine flux
(Ford~{\it et~al.}, 2001; Tinetti~{\it et~al.}, 2006; Williams and Gaidos,
2008). The presence of clouds, which cover an average of around $60
\%$ of the surface, increase Earth's albedo, as they dominate the
scattered light (Pall\'{e}~{\it et~al.}, 2008).  However, clouds on Earth are
generally tied to surface features and dissipate on the time scale of
days (Ford~{\it et~al.}, 2001; Pall\'{e}~{\it et~al.}, 2008), and therefore, they do
not vary greatly in the length of our earthshine
observations. Determining the weather conditions and surface features
of an extrasolar planet with no prior information is beyond the scope
of this paper.

Attempts to detect the proposed hydrocarbon oceans on Saturn's moon
Titan via specular reflection in 2003 and 2004 failed (West~{\it et~al.},
2005). However, the Cassini Titan Radar Mapper flyby in 2006 imaged
the surface of the satellite and detected what were determined to be
large northern liquid lakes (Stofan~{\it et~al.}, 2007).  Titan's
environment suggests that the lakes are methane, which replenish the
observed abundance in the atmosphere by evaporation (Sotin, 2007).  In
a similar way, the conditions of a suitable Earth-like planet will
dictate water based liquid oceans (Williams and Gaidos,
2008). Spectroscopy of the atmosphere of a planet (for example by TPF)
before measurement of a photometric light curve would likely resolve
any ambiguities about the composition of large bodies of liquid on a
planet's surface. McCullough (2006) recognized the ability of specular
reflection by oceans in determining the boundaries of continents on
Earth-like extrasolar planets and used physical models to demonstrate
the variation in polarization associated with oceans and
land. Numerical models of diurnal light curves were consistent with
models by Ford~{\it et~al.} (2001). The benefit of combining polarization
and flux observations of an extrasolar planet was also indicated by
Stam (2008). However, the phase angle dependence of specular
reflection competes with the need for increased flux for the detection
of extrasolar planets. We show that, even when the Earth is
illuminated at up to$ 70 \%$, the interruption of specular reflection by
the rotation of land into view of the Moon causes a large decrease in
earthshine flux, which indicates the presence of liquid water.

Previously, monitoring of earthshine has been performed for climate
change studies through the measurement of Earth's Bond albedo, with
estimates of $10 \%$ variability in albedo during a night of
observations that is averaged out in phase correction and data
combination (Pall\'{e}~{\it et~al.}, 2003; Qui~{\it et~al.},
2003). However, previously published light curves that show variation
in albedo are not specifically studied for correlation with continents
or liquid water, nor have they been compared to subsequent
measurements of the reflectivity of the same region, as is the purpose
of this paper (see e.g., Pall\'{e}~{\it et~al.}, 2003;
Monta\~{n}\'{e}s-Rodr\'{i}guez~{\it et~al.}, 2007).  Similarly,
spectroscopic studies use a single night of observations to quantify
the wavelength dependence of Earth's albedo
(Monta\~{n}\'{e}s-Rodr\'{i}guez~{\it et~al.}, 2005; Arnold, 2007).
Modeling studies of the spectra and light curves of Earth have
suggested that cloud cover will decrease the variability of
reflectivity (Ford~{\it et~al.}, 2001; Tinetti~{\it et~al.},
2006). However, recent work by Pall\'{e}~{\it et~al.} (2008) showed that the 24
hour period of rotation is still discernible from photometric
observations due to surface features and the stability of weather
systems on Earth.  In this paper, we present observations that show
variation in earthshine due to the rapidly changing surface properties
across the east coast of Africa. Our repeated results allow
statistical comparison between observations of the same area with
differing weather conditions. The goal of our repeated measurements
was to demonstrate experimentally the utility of a telescope such as
TPF with respect to characterization of extrasolar planets via
photometric monitoring.

\subsection*{Materials and Methods}

Through a series of observations of the waxing crescent phase of the
Moon in Cousins I and R filters (centered on 600 nm and 800 nm,
respectively), we monitored hourly variation in the reflectivity of
Earth centered along segments of constant latitude that intercept the
continental coast between the southern tip of Africa and the Middle
East.  For comparison, we also measured a light curve of the Moon as a
waning crescent to observe constant specular reflection from the South
Pacific Ocean (data set F2). Our observations are summarized in
Table~1 and are labeled as data sets C1--C5, F1--F2. The earthshine
light curves, normalized by the intensity of the initial observation,
are also presented along with the weather conditions for the
corresponding reflecting region of Earth's surface in Figure~2.
Observations were made from a site near Romsey, 100 km north of
metropolitan Melbourne, Australia $(37^{o}23'$~S~$144^{o}41'$~E), with use of
an $8 "$ Meade Schmidt-Cassegrain telescope as a mount for a $512 \times
512$ pixel charge-coupled device camera. The field of view is $20.8' \times
20.8'$. Images of the sunlit lunar crescent were taken
contemporaneously with earthshine images to calibrate for changing air
mass.  The same two regions of the Moon (defined by lunar coordinates)
were used to measure the moonshine and earthshine fluxes for a set of
observations on a single night.  Those earthshine regions for a waxing
crescent are shown in Figure~1 (panels {\bf A} and {\bf B}) and are marked by the
boxes.  Use of the same regions within a night allows comparison of
results, while limiting the systematic variations associated with the
relative reflectivity of different regions of the Moon.  When
possible, the same region of the Moon was used for all sets of
observation..  The time variation in absorption of flux was calculated
by fitting the crescent intensity with Beer's law of atmospheric
extinction.  The statistical uncertainty among crescent light curve
points does not fully account for the level of departure of the data
from the best fit.  We therefore introduced an additional scatter
(which might describe the presence of fluctuations in atmospheric
opacity around the smooth trend) at a level such that the best-fit
Beer's Law has a reduced $\chi^{2}$ of unity.  The earthshine light
curve was divided by the estimate of Beer's law, which yielded the
relative change in intrinsic earthshine flux.

Light from the bright crescent flares into view of the charge-coupled
device during earthshine exposures. This is particularly evident
during large lunar phases and is a known difficulty in the data
reduction for earthshine observations due to the telescope optics in
imaging the dim earthshine region adjacent to the bright
crescent. Previous studies have removed it by subtraction of a linear
extrapolation of the background, which decreases radially from the
center of the Moon (Qui~{\it et~al.}, 2003; Arnold, 2007).  In this work, we
introduced a new technique to remove the scattered light.  The
scattered light has low spatial curvature when compared to the sharp
lunar features (see Figure~1, panel {\bf B}).  After smoothing all images
from a night of observation on a scale that corresponds to the seeing
of the worst image (Gaussian across 6 pixels), we take the
pixel-to-pixel Laplacian of the earthshine flux in a region with a
highly variable surface, such as a crater.  In this region, the
scattered light is approximated as linearly changing with position.
Figure~1 shows examples of images with minimal and significant visible
scattered light (panels {\bf C}, {\bf D} respectively). The right hand panels show
the corresponding Laplacian along a line through the variable region
of the Moon (panels {\bf E}, {\bf F}).

Inspection of the Laplacian as a function of position reveals
fluctuations about zero across the entire observed area. The size of
these fluctuations is largest at positions coinciding with lunar
surface features, and smallest in the area that corresponds to the
sky. The utility of the method is demonstrated by panels {\bf B} and {\bf F},
which correspond to a particularly poor observation included here to
demonstrate that scattered light removal via the described Laplacian
technique is viable even in extreme cases. The Laplacian technique
leaves residuals about zero in regions of the sky, underneath the
prominent region of scattered light, that are as small as for the
quality of image in panel {\bf A}.

From the fact that the Laplacian has no trend to increase or decrease
away from minimal variation about zero in regions of significant
scattered light, as well as those with minimal scattered light, we
conclude that there is no residual contribution of scattered light to
the Laplacian.

On the other hand, features like craters show up as characteristic
double horned profiles. The amplitudes of these features are
proportional to the earthshine flux.  An important advantage of this
technique, which is designed specifically for our data analysis, is
that it requires no prior knowledge of the distribution of background
scattered light and is computationally simple.  We base our light
curve measurement on the maximum change in the Laplacian within a
finite region. By choosing a finite region (shown as the box in
Figure~1, panels {\bf A}, {\bf B}) rather than a point, we allow for
error in image alignment.  The regions are chosen to be the same for
each night of observations, but as different data sets are not
quantitatively compared via flux magnitude, an exact measure of the
scattered light is not required. This also removes the need to
quantify any change between nights in physical scale of the region
over which the Laplacian is taken. The earthshine region used in sets
C1--C5 and F1 is based on the Grimaldi crater. In set F2 it is based on
Mare Crisium.

\subsection*{Results}

In Figure~2, we show the results of our earthshine observations.  The
upper left panel (panel {\bf A}) is a map of the Earth with lines showing
the path taken by the central region of reflectance during each night
of observation.  For each night of observation, three panels are
shown. The upper panel shows the color variation with time, the
central panel shows the light curve, and the lower panel shows the
central reflecting region for the observation with local weather
information.  In sets C1--C5, we attribute the decrease in earthshine
flux of $10-20 \%$ to the interruption of specular reflection from the
Indian Ocean by the African coast rotating into view. In comparison,
the light curves in sets F1 and F2 show little variation, with
decreases of around $2\%$. This quiescence is due to the constant
reflective properties of the diffusive desert regions of northern
Africa (F1) and the constant specular reflection of the South Pacific
Ocean (F2).

For data sets C1--C5, there is variation in the decrease per hour
measured due to the differing local cloud cover, the latitude of the
central reflection (due to the time of year), and the length of time
for which the data is available. Data set C3 does not extend as far
west, and hence, specular reflection would not have been totally
interrupted. The remaining contribution to reflection by the Indian
Ocean at later times will depend on the wave distribution off the
coast (Williams and Gaidos, 2008). For data set C5, large local cloud
formations are present over the ocean; this type of cloud cover will
diffuse incident sunlight on the ocean and prevent specular
reflection.

\subsection*{Discussion}
 
To investigate the light curve trends quantitatively, we used a linear
function to parameterize the variation of earthshine intensity with
time.  This function has the form: $ I(t) = \alpha t + \beta$, where
the parameters $\alpha$ and $\beta$ correspond to the gradient and
intercept of a straight line for the intensity I(t), at time t, The
rates of decrease $[d(\Delta I/I(t_{0}))/dt$, where $\Delta I = \alpha
  (\Delta t) ]$ for each data set are presented in
Table~1. Uncertainties were determined from the projection of the
ellipse for which $\chi^{2}$ equals the minimum $\chi^{2}$ plus 1.
The data sets C1--C5 exhibit statistically significant decreases in
earthshine for observations spanning the African coast. Smaller
decreases are detected in sets C3 and C5 owing to oceanic cloud cover
and a lack of reflection points centered over continental Africa. The
data sets F1 and F2 exhibit small (consistent with zero) decreases in
earthshine due to the constant surface characteristics.  The short
time scale of the observations suggests that the cloud formations in
the Moon's line of sight, and their overall contribution to the
scattered light, will remain fairly stable. Results are compared
within a night of observations; therefore, the weather conditions are
most important when local cloud masks specular reflection in some
oceanic regions, for example, in set C5. As seen in earthshine
observations of the vegetation red edge, weather conditions dictate
the strength of the signal seen (Monta\~{n}\'{e}s-Rodr\'{i}guez~{\it
  et~al.}, 2006). However, in future long-term observations of
extrasolar planets with evident weather systems, the rotational period
may still be discernible by folding light curves (Pall\'{e}~{\it
  et~al.}, 2008). We require more data sets to determine whether there
is any dependence of our result on the phase angle of the Moon. We are
unable to observe the crescent Earth via earthshine and, therefore,
cannot measure the predicted increase in specular reflection due to
the grazing angle (Williams and Gaidos, 2008).

We note that the measured change in earthshine flux is not the result
of hourly variations in lunar phase. The intensity of reflected
sunlight increases by around $0.15 \%$ per hour at the lunar phase of
the observations (Qui~{\it et~al.}, 2003).  In addition, the increase
in lunar phase angle also corresponds to an increase in path length
through the atmosphere, which, for our observations, corresponds to
around a $1 \%$ per hour decrease in the intensity of earthshine.
Thus, the residual decrease in earthshine due to the Earth's phase as
seen from the Moon is below $1 \%$ during a typical night of
observations.

In addition to variable flux, our observations indicate associated
changes in color, which we interpret as being due to some combination
of the vegetation red-edge (Seager~{\it et~al.}, 2005; Hamdani~{\it
  et~al.}, 2006) and the increased reflectivity of deserts into the
infrared (which reddens the scattered light as it decreases in
brightness) (upper sub-panels in Figure~2). The results of a linear
parameterization are presented in Table 1, where statistically
significant reddening of the spectrum is measured for data sets C1,
C2, and C4. This effect is not seen in set F2, where the reflecting
surface is uniformly ocean. Moreover, there is a strong correlation
between the decline in flux and color change for each data set
(Figure~3).  Thus, observation of a coast line crossing in reflected
light from an extrasolar planet may be characterized not only by a
sharp decline of flux, but also by reddening of the reflected
spectrum.  These ideas are supported by modeling by Stam (2008), where
a cloudless planet shows more variation in the flux and polarization
in the near infrared (0.87 $\mu m$) than toward the blue (0.44 $\mu m$) due to
areas of vegetation compared to oceanic regions.  For a realistic
cloudy Earth model, the effect is decreased, as seen in earthshine
observations of the vegetation red edge (Arnold, 2007).  Further
observations of other coast lines and land masses are required to
determine whether the reddening of the spectrum is unique to our
observations of Africa or whether they may be used in characterizing
extrasolar planets as having continents and liquid oceans

\subsection*{Conclusions}

The consistent decrease in earthshine flux in repeated observations
corresponding to reflection from the African coast supports the idea
that photometric variability may be used by future space-based
missions to characterize terrestrial extrasolar planets, particularly
those that have significant fractions of their surfaces covered by
both seas and land masses. For example, surface features could be used
to determine the rotational period of Earth-like planets (Pall\'{e}~{\it et~al.},
 2008). We find the observed photometric variability associated
with a continental coast crossing to be substantially larger than the
measured spectral change due to vegetation's red edge (Arnold,
2007). In addition, unlike the photometric properties of land and
water, the spectral signature of vegetation might not be universal
(Seager~{\it et~al.}, 2003). Finally, variability is measured in broad bands
rather than spectra, which makes this terrestrial signature more
readily detectable. Further observations of the African coast and
other landmasses will help determine the detectability of a coast via
the reddening of the spectrum.  Determining the phase dependence of
the magnitude of the decrease, and more detailed analysis of the
effect of weather on Earth's variability, will require more
observations.  Spectropolarimetric modeling studies have suggested
that the combination of photometric flux and polarization will yield
the best characterization of surface types, therefore, further
earthshine observations are justified (Williams and Gaidos, 2008).

Our results highlight the importance of considering specular
reflection from oceans in the modeling and analysis of light curves
from Earth-like extrasolar planets, and suggest that it is a useful
tool in determining the presence of liquid water on a
planet. Increasingly sophisticated modeling studies of photometric
variability (Pall\'{e}~{\it et~al.}, 2008; Williams and Gaidos, 2008), which
account for the effects of viewing geometry, dynamic weather, and data
quality on light curve characterization (Ford~{\it et~al.}, 2001), are
therefore justified by empirical studies of the Earth as a model
extrasolar planet.

\vskip 0.2in

\bigskip
\bigskip
\noindent
{\bf Acknowledgments:}\\ We thank the anonymous referees for their detailed comments that enabled us to improve the clarity of the paper.
This work was supported in part by grants from the Australian Research Council. 
 SVL acknowledges the support of an Australian Postgraduate Award and a Postgraduate Overseas Research Experience Scholarship.

\bigskip
\bigskip
\noindent
{\bf Author Disclosure Statement:}\\ No competing financial interests exist.

\bigskip
\bigskip
\noindent
{\bf Abreviation:}\\ TPF, Terrestrial Planet Finder.

\pagebreak

\begin{figure*}[t]

\epsscale{0.75}  \plotone{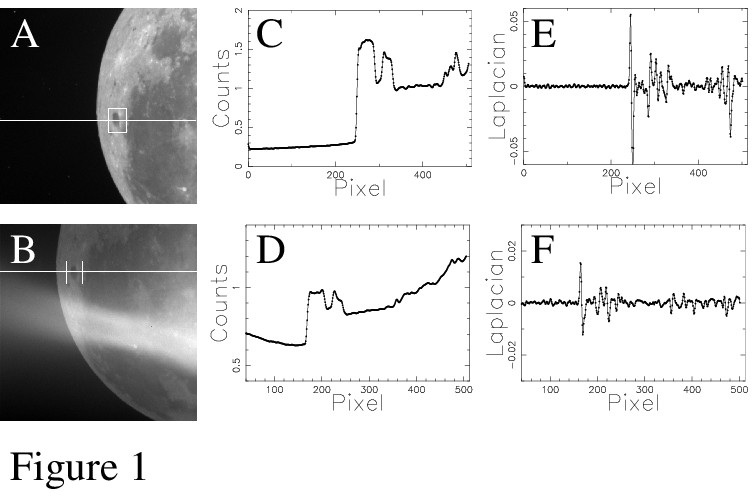}
\caption{\label{fig1} 
{\bf Example earthshine obserations.}
Panel {\bf A}: An example of an earthshine image with no visible scattered light. Panel {\bf B}: An example of a poor earthshine image with significant visible scattered light from telescope optics. Panels {\bf C,D}: Flux along a section through earthshine image, as indicated by the white line in panels {\bf A} and {\bf B}. Panels {\bf E,F}: Laplacian of image along same line. The maximum peaks correspond to the edge of the Moon.}
\end{figure*}

\pagebreak

\begin{figure*}[t]

\epsscale{1.0}  \plotone{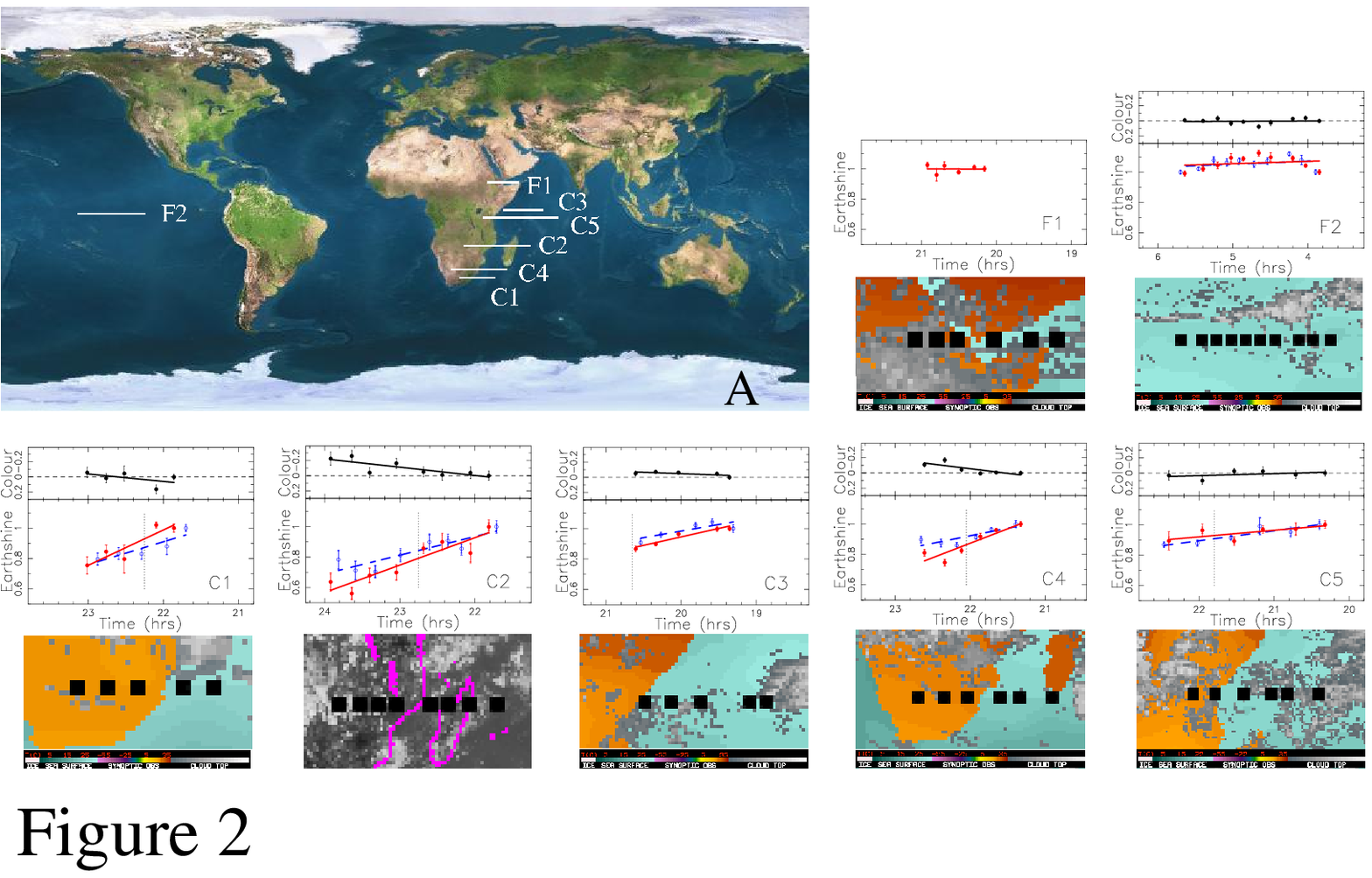}
\caption{\label{fig2} 
{\bf Earthshine light curves.} 
Panel {\bf A} shows the path of the central point of scattered light~(Earth and Moon Viewer; see http://www.fourmilab.ch/earthview/vplanet.html.). The remaining panels (F1, F2, C1, C2, C3, C4 and C5) show:  
{\it Middle sub-panel:} Light curves for the Cousins $R$ (red) and $I$ (blue) filters, with the coastal crossing time (dotted).
The smaller error bars show the statistical errors in photometry, the larger error bars include the intrinsic scatter in airmass
around Beer's law.
{\it Top sub-panel:} Color change ($R-I$) with the corresponding linear fit (solid). 
{\it Bottom sub-panel:} Weather details for the region of reflected light with
black boxes representing the center of reflection.
 }
\end{figure*}

\pagebreak

\begin{figure*}[t]

\epsscale{0.5}  \plotone{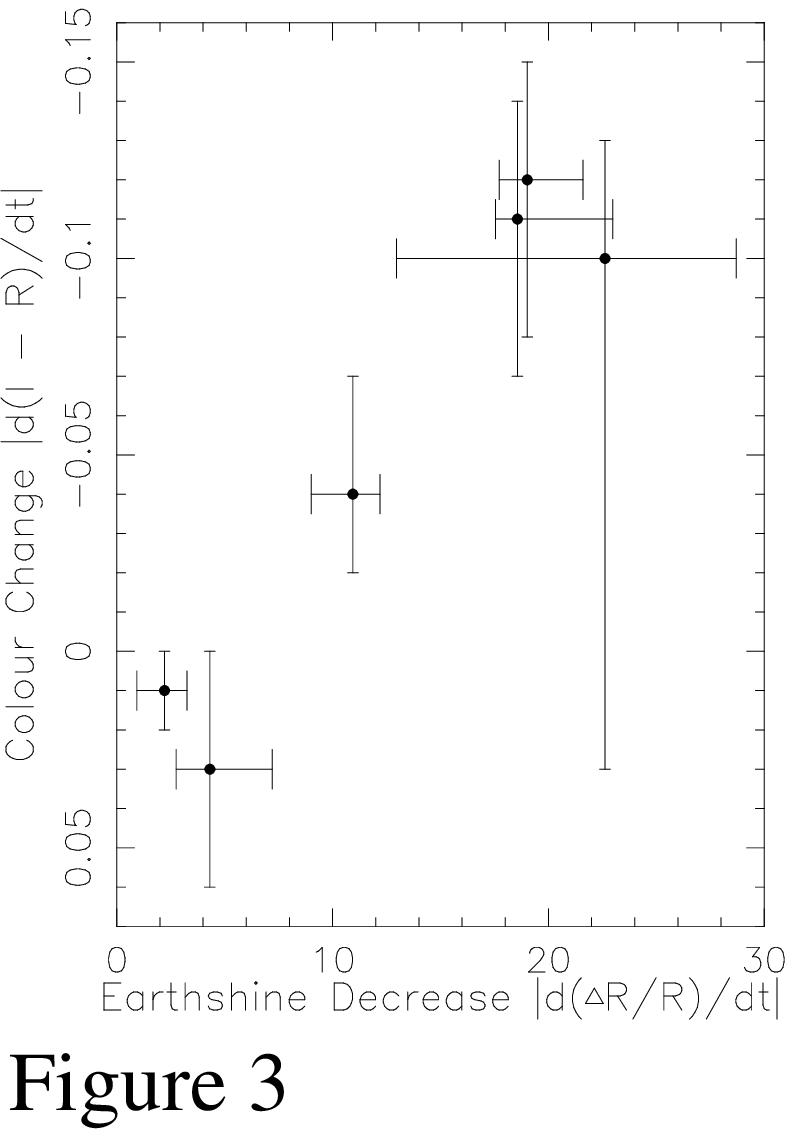}
\caption{\label{fig3} 
{\bf Correlation of flux change and color change.}
The magnitude of color change per hour 
for data sets C1--C5 and F2 are plotted versus the size of the decline
in earthshine per hour in the R filter (solid). 
Error bars are quoted as 1$\sigma$ throughout this {\em Paper}.}
\end{figure*}

\pagebreak

\begin{table} [b]
\centering
\begin{tabular}{|c|c|c|c|c|c|c|c|}
\hline

{\footnotesize \bf Set}&{\footnotesize \bf Date}&{\footnotesize \bf Phase}&{\footnotesize \bf Filter}&{\footnotesize \bf Time $\Delta t$}&{\footnotesize \bf $\chi^{2}_{R}$}&{\footnotesize \bf $d(\Delta I/I_{0})/dt$}&{\footnotesize \bf $d(I-R)/dt$}\\
\hline
\hline

{\footnotesize C1}&{\footnotesize 12/04/05}&{\footnotesize 8\% waxing}&{\footnotesize I}&{\footnotesize 1.18 hrs}&{\footnotesize 1.74}&{\footnotesize -15.71$\pm^{7.82}_{2.43}$ \%}&{\footnotesize -0.10$\pm^{0.13}_{0.03}$ \%} \\ [0.11 cm]

 &&&{\footnotesize R}&{\footnotesize 1.16 hrs}&{\footnotesize 3.15}&{\footnotesize -22.62$\pm^{6.08}_{9.66}$ \%}& \\ [0.11 cm]
\hline

{\footnotesize C2}&{\footnotesize 01/05/06}&{\footnotesize 31\% waxing}&{\footnotesize I}&{\footnotesize 2.11 hrs}&{\footnotesize 0.99}&{\footnotesize -12.54$\pm^{2.99}_{1.82}$ \%}&{\footnotesize -0.11$\pm^{0.05}_{0.03}$ \%}  \\ [0.11 cm]

&&&{\footnotesize R}&{\footnotesize 2.11 hrs}&{\footnotesize 1.76}&{\footnotesize -18.56$\pm^{4.42}_{1.01}$ \% }&\\ [0.11 cm]
\hline

{\footnotesize C3}&{\footnotesize 08/28/06}&{\footnotesize 16\% waxing}&{\footnotesize I}&{\footnotesize 1.23 hrs}&{\footnotesize 2.52}&{\footnotesize  -8.28$\pm^{2.73}_{1.77}$ \% }&{\footnotesize -0.04$\pm^{0.02}_{0.03}$ \%}\\ [0.11 cm]

&&&{\footnotesize R}&{\footnotesize 1.25 hrs}&{\footnotesize 2.37}&{\footnotesize -10.94$\pm^{1.26}_{1.93}$ \% }& \\ [0.11 cm]
\hline

{\footnotesize C4}&{\footnotesize 11/24/06}&{\footnotesize 10\% waxing}&{\footnotesize I}&{\footnotesize 1.26 hrs}&{\footnotesize 3.52}&{\footnotesize -10.10$\pm^{3.09}_{1.17}$ \%} &{\footnotesize -0.12$\pm^{0.04}_{0.03}$ \%}\\ [0.11 cm]

&&&{\footnotesize R}&{\footnotesize 1.28 hrs}&{\footnotesize 4.91}&{\footnotesize -19.02$\pm^{2.58}_{1.30}$ \% } &\\ [0.11 cm]
\hline

{\footnotesize C5}&{\footnotesize 08/19/07}&{\footnotesize 32\% waxing}&{\footnotesize I}&{\footnotesize 2.07 hrs}&{\footnotesize 0.47}&{\footnotesize -6.61$\pm^{1.46}_{1.97}$ \% }&{\footnotesize 0.03$\pm^{0.03}_{0.04}$ \%} \\ [0.11 cm]

&&&{\footnotesize R}&{\footnotesize 2.08 hrs}&{\footnotesize 1.04}&{\footnotesize -4.38$\pm^{2.89}_{1.56}$ \% }& \\ [0.11 cm]

\hline
\hline

{\footnotesize F1}&{\footnotesize 07/29/06}&{\footnotesize 14\% waxing}&{\footnotesize I}&-&-&-&-\\ [0.11 cm]

&&&{\footnotesize R}&{\footnotesize 0.76 hrs}&{\footnotesize 1.78}&{\footnotesize 0.0003$\pm^{0.002}_{0.006}$\% }&\\ [0.11 cm]
\hline

{\footnotesize F2}&{\footnotesize 04/14/07}&{\footnotesize 16\% waning}&{\footnotesize I}&{\footnotesize 1.80 hrs}&{\footnotesize 6.1}&{\footnotesize -1.88$\pm^{3.08}_{0.25}$ \%} &{\footnotesize 0.001$\pm^{0.004}_{0.014}$ \%}  \\ [0.11 cm]

&&&{\footnotesize R}&{\footnotesize 1.79 hrs}&{\footnotesize 6.17}&{\footnotesize -2.21$\pm^{1.04}_{1.28}$ \% }& \\ [0.11 cm]
\hline

\end{tabular}
\label{oblog}
 \caption{{\bf Observational log and results from analysis of a linear fit to light curves.}
The quantity $\chi^{2}_{R}$ is the reduced $\chi^{2}$ value of the fit with scatter around Beer's law.
The percentage decrease is the decrease in earthshine intensity per hour for the linear fit. The far right column is the magnitude of the color change.
}
\end{table}

\end{document}